\documentclass[aps,prb,floats,showpacs,twocolumn]{revtex4-1}
\usepackage{graphicx,color}
\begin{document}
\title{Temperature-dependent resistivity of suspended graphene}
\author{Eros Mariani$^1$ and Felix von Oppen$^2$}
\affiliation{$^1$ Centre for Graphene Science, School of Physics, University of Exeter, Stocker Rd., EX4 4QL Exeter, UK\\
$^2$Dahlem Center for Complex Quantum Systems and Fachbereich Physik, Freie Universit\"at Berlin, 14195 Berlin, Germany}
\date{\today}
\begin{abstract}
In this paper we investigate the electron-phonon contribution to the resistivity of suspended single layer graphene. In-plane as well as flexural phonons are addressed in different temperature regimes. 
We focus on the intrinsic electron-phonon coupling due to the interaction of electrons with elastic deformations in the graphene membrane. The competition between screened deformation potential vs fictitious gauge field coupling is discussed, together with the role of tension in the suspended flake.
In the absence of tension, flexural phonons dominate the phonon contribution to the resistivity at any temperature $T$ with a  $T^{5/2}_{}$ and $T^{2}_{}$ dependence at low and high temperatures, respectively. Sample-specific tension suppresses the contribution due to flexural phonons, yielding a linear temperature dependence due to in-plane modes. 
We compare our results with recent experiments.
%compatible with recent experimental observations. 
%The flexural phonon contribution could be detected in samples with reduced tension and is likely to be involved in the density dependence observed in recent measurements on suspended graphene devices.
%Finally, in order to address the unexpected density dependence shown by suspended graphene devices, we discuss the role of screening, non-degenerate Fermi gas corrections as well as the capacitive coupling with the back-gate. The latter induces a linear coupling for flexural modes resulting in a density dependent resistivity which could be relevant to address the experimental data.

\end{abstract}
 \maketitle
%%%%%%%%%%%%%%%%%%%%%%%%%%%%%%%%%%%%%%%%%%%%%%%

\section{Introduction}
In recent years the discovery of graphene, \cite{Geim,Kim} a monolayer of carbon atoms arranged in a honeycomb lattice, stimulated an unprecedented interest from physicists belonging to different communities. Electrons in graphene behave as massless Dirac fermions, as confirmed by their peculiar Quantum Hall Effect. \cite{Geim,Kim} In parallel, graphene represents the only existing two-dimensional (2D) conducting membrane embedded in 3D space. This feature became most prominent with the experimental realisation of suspended graphene devices. \cite{Meyer,McEuen} The latter allow for the investigation of the intrinsic properties of the material, unperturbed by the presence of a substrate. Thus, suspended graphene offers the unique possibility of exploring a system involving at once features of quantum-electrodynamics as well as of hard and soft condensed matter physics. \cite{KimNeto}

For a long time the very existence of 2D membranes was thought to be impossible \cite{Peierls,Landau} due to their tendency towards spontaneous crumpling. Indeed, at harmonic level, the elastic theory of flat 2D membranes yields divergent fluctuations of the angles of the normal vectors to the membrane, due to thermal fluctuations of flexural (out-of-plane) phonons. It was later understood that the non-linear coupling between stretching and bending energies hardens the bending stiffness at long wavelengths and stabilises the flat phase. \cite{Nelson,NelsonCrumpling,Mariani}

It was soon realised that mechanical deformations of graphene sheets affect the electronic properties by inducing coexistent scalar and (fictitious or synthetic) gauge fields in the effective low-energy Dirac Hamiltonian. \cite{Mahan,Suzuura,Vozmediano} Recently this issue became of special relevance and stimulated the emergence of the so-called strain-engineering community, aimed at controlling the electronic properties of graphene by suitably engineering the deformations. \cite{Pereira,PacoNat}

Suspended membranes show flexural deformations which are typically extremely soft compared to in-plane ones. It may be tempting to conclude that the former should dominate the low-energy electromechanical properties of suspended graphene due to their large density of states. This effect is balanced by the strength of the electron-phonon coupling in graphene. While in-plane phonons show a conventional linear coupling, the intrinsic electron-phonon coupling of the flexural modes is quadratic. \cite{Suzuura,Mariani} This feature is protected by the reflection symmetry with respect to the plane, as the effect of out-of-plane modes cannot depend on the sign of the deformations. Thus, we find an interesting competition between hard-to-excite but strongly-coupled in-plane phonons and soft but weakly-coupled flexural ones. 

From an experimental point of view, this competition can be addressed, e.g., by transport measurements analyzing the temperature dependent resistivity $\Delta \rho$.  Indeed, at high enough electron concentrations and for not too low temperatures (where quantum effects become prominent \cite{Kozikov}), $\Delta \rho$ is ascribed to electron-phonon scattering.
Recent measurements \cite{Morozov,Fuhrer,Bolotin} show a $\Delta \rho$ scaling linearly with temperature $T$. This behaviour is compatible with the expectations for longitudinal in-plane phonons \cite{Harrison,Hwang08} at $T>T^{(l)}_{{\rm BG}}$ (with $T^{(l)}_{{\rm BG}}$ the related Bloch-Gr\"uneisen temperature) as if the flexural phonon contribution were irrelevant.
In addition, non-suspended samples \cite{Fuhrer} show a density independent $\Delta \rho$ while suspended graphene \cite{Bolotin} exhibit a density dependence which has not been addressed so far. 
%
%From a theoretical point of view, in-plane phonons are expected to yield no density dependence in the resistivity and cannot account for the behaviour of suspended samples. While it seems natural that the presence of the substrate suppresses flexural-fluctuations, in suspended devices they could still play a relevant role and give rise to new interesting physics, possibly involving a residual density dependence.
%
%In addition, non-suspended samples \cite{Fuhrer} show a density independent $\Delta \rho$ in agreement with the theoretical expectations for in-plane phonons alone, suspended graphene \cite{Bolotin} shows a density dependence which has not been addressed so far. While it seems natural that the presence of the substrate suppresses flexural-fluctuations, in suspended devices they could still play a relevant role and give rise to new interesting physics, possibly involving a residual density dependence.

In a previous paper \cite{Mariani} we investigated the low-temperature dependence of the resistivity of suspended graphene in the absence of tension and found that flexural modes should actually dominate over the in-plane ones yielding a $T^{5/2} \log T$ dependence.
Motivated by the experimental puzzles above, here we further analyse the different contributions of in-plane vs flexural phonons to the temperature dependent resistivity by exploring the high-temperature regime and addressing further issues related to non-universal factors, like contact-induced tension in the membrane. 
In particular we investigate whether flexural modes keep dominating the $T$-dependence of the resistivity also at high temperatures and how tension affects their contribution with respect to in-plane phonons. We discuss the dependence of the resistivity on temperature as well as on electron density in light of the recent experimental findings.
%We find that, for the electron densities achievable in suspended graphene, flexural phonons should dominate in-plane ones {\em at any temperature in the absence of tension}. In particular, the flexural resistivity should show a low-temperature $T^{5/2}$ scaling crossing over to $T^2$ for $T\gg T_{{\rm BG}}^{({\rm flex})}$. A sample-specific tension induced by the contacts yields a stiffening of the flexural dispersion (much like in guitar-strings tuning) corresponding to a suppressed density of states, while the up-down symmetry protects their weak quadratic coupling. As a result, tension suppresses the flexural contribution which shows a low $T$ resistivity scaling as $T^7$, crossing over to $T^2$ for $T\gg T_{{\rm BG}}^{({\rm flex})}$. 
%We conclude that it is due to the non-universal tension-induced suppression of flexural modes that experiments seem to show only the in-plane contribution. If the tension in the sample could be lowered, the flexural phonons contribution should become dominant again. 

The structure of the paper is as follows: In Sec.\ \ref{sec:Basics} we introduce the basic description of the electronic and phononic properties of graphene to be employed in the rest of the manuscript. In Sec.\ \ref{sec:Resistivity} we discuss the in-plane vs flexural phonon contributions to the resistivity within a Boltzmann approach. We deduce the temperature and density dependence of the resistivity $\Delta\rho$ in the different regimes where acoustic phonons are relevant. We also comment on the effect of screening in the density dependence of the electron-phonon resistivity. Finally, we conclude in Sec.\ \ref{sec:Conclusions}.

\section{Electrons and phonons in graphene}
\label{sec:Basics}

\subsection{Electronic properties} 
In this paper we consider the scattering between electrons in graphene and long-wavelength phonon modes. We can thus treat the two Dirac cones independently and focus on the effective low energy Dirac Hamiltonian \cite{Wallace,DiVincenzo}
\begin{equation}
\label{HDirac}
H=\hbar v\,\mbox{\boldmath$\sigma$}\cdot\mathbf{k}
\end{equation}
where $v\simeq 10^{6}_{}\,{\rm m\cdot s_{}^{-1}}$ denotes the Fermi velocity and the 2D wavevector $\mathbf{k}$ is measured from the relevant Dirac point. 
%$\pm {\bf k}_{D}^{}$, with $\mathbf{k}_{D}^{}=2\pi/(3\sqrt{3}a)\, (\sqrt{3},1)$ and $a=1.42\, \AA$ the bond length. 
The Hamiltonian in Eq.\ (\ref{HDirac}) acts on two-component spinors $(u_{A,\mathbf{k}}^{},u_{B,\mathbf{k}}^{})$ of Bloch amplitudes in the space spanned by the two inequivalent honeycomb sublattices ($A-B$). The components of the vector 
$\mbox{\boldmath$\sigma$}$ are the Pauli matrices in the sublattice space. The real spin degree of freedom does not play a role here and will be ignored. 
The electronic spinor eigenstates $|{\bf k}\rangle$ with chirality $s= \mbox{\boldmath$\sigma$}\cdot {\bf k}/|{\bf k}|=\pm 1$ have energy $\epsilon_{{\bf k}}^{}= s\hbar v |{\bf k}|$. Their real space representation is $\langle {\bf r}|{\bf k}\rangle =1/\sqrt{2} (1, s\exp [i\phi_{{\bf k}}^{}]) \exp [i{\bf k}\cdot {\bf r}]$, with ${\bf r}$ the position vector in the 2D plane and $\phi_{{\bf k}}^{}$ the angle of the vector ${\bf k}$ with respect to the $x$-axis fixed along the armchair direction of the graphene lattice. 
%on graphene and we choose normalisation to the unit area.

\subsection{Phononic properties} 
The mechanical distortions of graphene are described by the vector field ${\bf u}({\bf r})$ and by the scalar field $h({\bf r})$ associated with in-plane and flexural (out-of-plane) deformations, respectively. The physics of the mechanical distortions is captured by the elastic Lagrangian density
\begin{eqnarray}
    &&{\cal L} = {\cal L}_{{\rm stretch}}^{}+{\cal L}_{{\rm bend}}^{}\nonumber \\
    &&{\cal L}_{{\rm stretch}}^{}=\frac{\rho_0}{2} \dot{\bf u}^2 - \mu
    u^2_{ij}-\frac{1}{2} \lambda u^2_{kk} \\
    &&{\cal L}_{{\rm bend}}^{}=\frac{\rho_0}{2}\dot h^2-\frac{1}{2}\kappa_{}^{} (\nabla^2 h)^2-\frac{1}{2}\gamma (\nabla h)^2\nonumber  
    \label{elastic}
\end{eqnarray}
with contributions coming from stretching and bending energies.
Here $\rho_0$ is the mass density of graphene and 
\begin{equation}\label{Strain}
u_{ij}=\frac{1}{2}[\partial_i u_j + \partial_j u_i + (\partial_i h) (\partial_j h)] 
\end{equation}
the strain tensor, \cite{LandauBook} with $i,j\in \{ x,y\}$. The Lam\'e coefficients $\lambda$ and $\mu$ characterise the in-plane rigidity of the lattice, while $\kappa_{}^{}$ is the bending stiffness and $\gamma$ is a sample-specific coefficient associated with tension induced by the edges of the sample. \cite{Parameters} Notice that the $\gamma (\nabla h)^2$ term breaks rotational symmetry which would be obeyed in the absence of tension. 

In the harmonic approximation the Lagrangian above yields two in-plane phonon modes (longitudinal $l$ and transverse $t$) and one flexural branch ($h$) with dispersions
\begin{eqnarray}\label{Dispersions}
&&\omega^{(l)}_{\mathbf{q}}=v^{(l)}q \nonumber\\
&&\omega^{(t)}_{\mathbf{q}}=v^{(t)}q \\
&&\omega^{(h)}_{\mathbf{q}}=\sqrt{\left(\gamma q^2+\kappa q^4\right)/\rho_{0}^{}} \nonumber
\end{eqnarray}
and group velocities $v^{(l)}=\left[\left(2\mu +\lambda\right)/\rho_0\right]^{1/2}\simeq 2\cdot 10^4_{}\, {\rm m}/{\rm s}$ and $v^{(t)}=\left[\mu/\rho_0\right]^{1/2} \simeq 1.3\cdot 10^4_{}\, {\rm m}/{\rm s} $. 
At long wavelengths (with respect to $a=1.42\, {\rm \AA}$, the graphene lattice spacing), in-plane phonons have a linear dispersion as Goldstone modes associated with the breaking of translational invariance in the plane. In contrast, flexural modes have a quadratic dispersion in the absence of tension. Tension introduces a new wavevector scale $q^{}_{*}=(\gamma /\kappa )^{1/2}_{}$ discriminating a tension-induced linear dispersion at low momenta from the quadratic dependence at $q>q^{}_{*}$
\begin{eqnarray}\label{FlexLimits}
&&\omega^{(h)}_{\mathbf{q}}\simeq \alpha q\quad\quad {\rm for}\;  q\ll q^{}_{*}\nonumber\\
&&\omega^{(h)}_{\mathbf{q}}\simeq \beta q^{2}_{}\quad\;\; {\rm for}\;  q\gg q^{}_{*}\nonumber
\end{eqnarray}
with $\alpha =(\gamma /\rho_{0}^{})^{1/2}_{}$ and $\beta =(\kappa /\rho_{0}^{})^{1/2}_{}$.
Correspondingly, the density of states (DOS) for in-plane modes is linear in energy $\epsilon$ while that for flexural phonons is linear in energy up to $\hbar\omega^{(h)}_{q_*}$ and independent of energy for $\epsilon>\hbar\omega^{(h)}_{q_*}$. As we will show, the DOS reduction induced by tension is the basic mechanism suppressing the contribution of flexural phonons relative to the tension-independent in-plane ones. 

In the following it will be useful to Fourier transform the phononic displacements $x^{(\nu )}_{}({\bf r})$ (with $\nu=l,t,h$) as $x^{(\nu )}_{}({\bf r})=\sum_{{\bf q}}^{}x^{(\nu )}_{{\bf q}}\exp [i{\bf q}\cdot {\bf r}]$. The normal modes can then be expressed as $x^{(\nu )}_{{\bf q}}=\xi _{{\bf q}}^{(\nu )}(a_{{\bf q}}^{(\nu )}+a_{{-\bf q}}^{(\nu )\dagger})$ with $\xi _{{\bf q}}^{(\nu )}=[\hbar /2M\omega^{(\nu )}_{{\bf q}}]^{1/2}_{}$ the oscillator length, $a_{{\bf q}}^{(\nu )}$ the annihilation operator for the mode $\nu$ at wavenumber ${\bf q}$ and $M$ the total oscillator mass per unit area.

\subsection{Electron-phonon coupling} 
In this paper we focus on intrinsic coupling mechanisms between electrons and phonons due to the effect of deformations on the electronic Hamiltonian. 
Other sample-specific mechanisms exist, e.g. due to the capacitive coupling between a back-gate and electrons in suspended graphene or due to buckling of the membrane (which breaks the reflection symmetry of the membrane), but these are beyond the scope of the present work. 
%We will briefly comment on those at the end of the paper but the analysis of the complete scenario of different possible couplings will be deferred to a future publication.
%In transport experiments through suspended carbon-nanotubes the capacitive coupling has been shown to give a small contribution with respect to the intrinsic one. The role of these additional effects will be analysed in a future publication.

The intrinsic coupling between electrons and deformations is related to the variation of areas and lengths induced in the membrane by specific distortions. As the variations of length or area are described by the components of the strain tensor, \cite{LandauBook} by examining its form in (\ref{Strain}) one readily concludes that in-plane phonons have a linear coupling to electrons while flexural phonons have a quadratic one, as long as the reflection symmetry with respect to the plane is not broken. 
We point out that the breaking of reflection symmetry (e.g. via capacitive coupling or via buckling) would yield a non-universal linear coupling for flexural modes. In the presence of tension this would effectively result in a sample-specific contribution to the resistivity with identical parametric dependences as for in-plane phonons.

In more detail, the representation of the electron-phonon coupling in the electronic Dirac description of a single valley is given by \cite{Mariani,Mahan,Suzuura}
\begin{equation}
 \label{Velph}
  V_{e-ph}^{}=  \left(\begin{array}{cc}g_{1}^{}(u_{xx}+u_{yy}) & g_
{2}^{}f^{*}_{}[u_{ij}^{}]
\\g_{2}^{}f[u_{ij}^{}] & g_{1}^{}(u_{xx}+u_{yy})\end{array}\right)\; ,
\end{equation}
with $f[u_{ij}^{}]=2u_{xy}+i (u_{xx}-u_{yy})$. The diagonal part of the coupling constitutes a scalar deformation potential originating from local area variations. The corresponding bare coupling constant has been estimated to be \cite{Suzuura} $g_{1}^{}\simeq 20-30\, {\rm eV}$.
In addition, distortions which induce no variation of local areas (e.g. pure shear modes) would still couple to electrons via the induced bond-length modulations. These affect the hopping amplitudes between neighbouring carbon atoms and induce the off-diagonal terms of Eq.\ (\ref{Velph}) corresponding to a fictitious (or synthetic) gauge field in the Dirac equation. 
The corresponding coupling constant has been estimated to be \cite{Suzuura} $g_{2}^{}\simeq1.5\, {\rm eV}$ which is about an order of magnitude weaker than the deformation potential coupling.
In the following, we will assume that these estimates are at least roughly correct. However, we emphasize that it would be straightforward to adapt our results to situations where this inequality no longer holds.

It should be noticed however that the deformation potential, contrary to the gauge field coupling, is affected by electronic screening \cite{Hwang07,FelixPacoEros} which reduces the coupling constant $g_1^{}$. Indeed, if ${\bf Q}$ is the wavevector transferred in the electron-phonon coupling, Thomas-Fermi screening of the deformation potential yields the coupling constant 
\begin{equation}\label{Screening}
g_{1}^{({\rm sc})}(Q)=g_{1}^{}\frac{1}{1+{\cal V}_{{\bf Q}}^{}\Pi_{{\bf Q}}^{}}= g_{1}^{}\frac{Q}{Q+Q_{{\rm TF}}^{}}
\end{equation}
with $Q=|{\bf Q}|$, ${\cal V}_{{\bf Q}}^{}=2\pi e^2_{}/Q$ the 2D Coulomb interaction and $\Pi_{{\bf Q}}^{}$ the fermionic polarization. 
The Thomas-Fermi screening wavevector $Q_{{\rm TF}}^{}=2\pi e^2 \nu_{{\rm F}}^{}=\eta \alpha_{{\rm f}}^{} k_{{\rm F}}^{}$ is expressed via the fine structure constant of graphene $\alpha_{{\rm f}}^{} =e^{2}_{}/\hbar v\simeq 2$ and the electronic DOS at the Fermi level $\nu_{{\rm F}}^{}=\eta k_{{\rm F}}^{}/2\pi\hbar v$, with $\eta =4$ due to the spin and valley degeneracy. 
For small wavevectors $Q\ll Q_{{\rm TF}}^{}$ the deformation potential coupling is thus strongly suppressed. 
We can identify a crossover wavevector $Q_{{\rm GD}}^{}=Q_{{\rm TF}}^{}g_{2}^{}/(g_{1}^{}-g_{2}^{})\ll 2k_{{\rm F}}^{}$ below which the gauge field coupling dominates over the deformation potential.

In summary, the longitudinal in-plane and flexural modes, as they induce local variations of area, couple to electrons via both the deformation potential and the gauge field mechanisms, while transverse in-plane phonons involve pure shear and couple only via the gauge field. At wavevector ${\bf Q}$, these phonon modes are characterized by the coupling matrices
\begin{eqnarray}
 \label{Couplings}
&&V_{e-ph}^{(l)}= w_{{\bf Q}}^{(l)}\left(a_{{\bf Q}}^{(l)}+a_{-{\bf Q}}^{(l)\dagger}\right)\nonumber \\
&&V_{e-ph}^{(t)}= w_{{\bf Q}}^{(t)}\left(a_{{\bf Q}}^{(t)}+a_{-{\bf Q}}^{(t)\dagger}\right)\\
&&V_{e-ph}^{(h)}= w_{{\bf q}_1^{},{\bf q}_2^{}}^{(h)}\left(a_{{\bf q}_1^{}}^{(h)}+a_{-{\bf q}_1^{}}^{(h)\dagger}\right)\left(a_{{\bf q}_2^{}}^{(h)}+a_{-{\bf q}_2^{}}^{(h)\dagger}\right)\; ,\nonumber 
\end{eqnarray}
with
\begin{eqnarray}
 \label{W}
&&w_{{\bf Q}}^{(l)}= iQ\xi_{Q}^{(l)} \left(\begin{array}{cc}g_{1}^{({\rm sc})}(Q) & -ig_{2}^{}e^{i2\Phi}_{}
\\ig_{2}^{}e^{-i2\Phi}_{} & g_{1}^{({\rm sc})}(Q)\end{array}\right)\nonumber \\
&&w_{{\bf Q}}^{(t)}= iQ\xi_{Q}^{(t)} \left(\begin{array}{cc}0 & g_{2}^{}e^{i2\Phi}_{}
\\g_{2}^{}e^{-i2\Phi}_{} & 0\end{array}\right) \\
&&w_{{\bf q}_1^{},{\bf q}_2^{}}^{(h)}= -\frac{q_{1}^{}q^{}_{2}}{2}\xi_{q_{1}^{}}^{(h)}\xi_{q_{2}^{}}^{(h)} \times\nonumber \\
&&\quad\quad\quad\times\left(\begin{array}{cc}g_{1}^{({\rm sc})}(Q)\cos \phi & -ig_{2}^{}e^{i\psi}_{}
\\ig_{2}^{}e^{-i\psi}_{} & g_{1}^{({\rm sc})}(Q)\cos \phi\end{array}\right)\nonumber 
\end{eqnarray}
in the Dirac description.
For flexural modes ${\bf Q}={\bf q}_1+{\bf q}_2$, with ${\bf q}_1$ and ${\bf q}_2$ the wavevectors of the two flexural phonons. In the matrices above $\Phi$, $\phi_1$ and $\phi_2$ are the angles of the vectors ${\bf Q}$, ${\bf q}_1$ and ${\bf q}_2$ with respect to the $x$-axis, respectively, and we defined $\phi=\phi_1-\phi_2$ and $\psi=\phi_1+\phi_2$.

\section{Contribution to the resistivity from electron-phonon scattering}
\label{sec:Resistivity}

We now address the electron-phonon contribution to the resistivity of suspended graphene. It is believed that this yields the dominant temperature dependence of the resistivity for the regime of high temperatures  involved in recent experiments \cite{Fuhrer,Bolotin} ($50\, {\rm K}<T<300\, {\rm K}$).
Further mechanisms contributing to the temperature dependence of the resistivity discussed in the literature focus on the ballistic regime, \cite{Trauzettel} on the $T$-dependent screening of Coulomb impurities,  \cite{Hwang09} or on the combined effects of electron-electron interactions and atomic-scale impurities. \cite{Cheianov} 

We focus on the high electron density regime where the Fermi wavevector $k_{{\rm F}}^{}$ is larger than the inverse mean free path due to disorder and electron-phonon scattering. In this regime a quasiclassical Boltzmann approach to transport can be employed.
The linearized Boltzmann equation for the electronic distribution function $f_{{\bf k}}^{}$ in presence of a constant electric field ${\bf E}$ has the standard form \cite{Landau10}
\begin{equation}
\label{Boltzmann}
-e{\bf E}\cdot {\bf v}\, \frac{\partial f_{{\bf k}}^{0}}{\partial \epsilon_{{\bf k}}^{}}=C(\varphi_{{\bf k}}^{})\; ,
\end{equation}
with $-e$ the electron charge and ${\bf v}=\partial \epsilon_{{\bf k}}^{}/\partial (\hbar{\bf k})$ the electron velocity.
Here we assumed the stationary condition $\partial f_{{\bf k}}^{}/\partial t =0$ as well as spatial uniformity, $\partial f_{{\bf k}}^{}/\partial {\bf r} =0$, and we denoted by $f_{{\bf k}}^{0}=(\exp [(\epsilon_{{\bf k}}^{}-\mu)/k_{{\rm B}}^{}T]+1)^{-1}_{}$ the equilibrium Fermi-Dirac distribution function, with $\mu$ the chemical potential. The collision integral on the right hand side of Eq.\ (\ref{Boltzmann}) depends linearly on the deviation from the equilibrium distribution, expressed through the function $\varphi_{{\bf k}}^{}$ as
\begin{equation}
\label{Varphi}
f_{{\bf k}}^{}-f_{{\bf k}}^{0}=-\frac{\partial f_{{\bf k}}^{0}}{\partial \epsilon_{{\bf k}}^{}}\, \varphi_{{\bf k}}^{}\; .
\end{equation}
In the relaxation time approximation we can express the collision integral as
\begin{equation}
\label{Relax}
C(\varphi_{{\bf k}}^{})=-\frac{f_{{\bf k}}^{}-f_{{\bf k}}^{0}}{\tau_{{\bf k}}^{}}=\frac{\varphi_{{\bf k}}^{}}{\tau_{{\bf k}}^{}}\,\frac{\partial f_{{\bf k}}^{0}}{\partial \epsilon_{{\bf k}}^{}}
\end{equation}
in terms of the transport scattering time $\tau_{{\bf k}}^{}$. 
%In the next sections we will discuss in detail the linearized collision integral due to electron-phonon scattering and deduce the corresponding $\tau_{{\bf k}}^{}$. 
By direct comparison of Eqs.\ (\ref{Boltzmann}) and (\ref{Relax}) we obtain the solution of the Boltzmann equation as
\begin{equation}
\label{BoltzSol}
\varphi_{{\bf k}}^{}=-e{\bf E}\cdot {\bf v}\, \tau_{{\bf k}}^{}\propto \cos \zeta
\end{equation}
with $\zeta$ the angle between ${\bf E}$ and ${\bf v}$. Notice that in graphene the modulus of the velocity is independent of wavevector and ${\bf v}\, \| \,\hat{{\bf k}}$.
The solution to the linearized Boltzmann equation yields the current density
\begin{equation}
\label{Current}
{\bf j}=-\eta e\sum_{{\bf k}}^{}{\bf v}\left(f_{{\bf k}}^{}-f_{{\bf k}}^{0}\right)=-\eta e^2_{}\sum_{{\bf k}}^{}\frac{\partial f_{{\bf k}}^{0}}{\partial \epsilon_{{\bf k}}^{}}\, {\bf v}({\bf E}\cdot {\bf v})\, \tau_{{\bf k}}^{}\; .\nonumber
\end{equation}
In the regime $k_{{\rm B}}^{}T\ll \epsilon_{{\rm F}}^{}$ (with $\epsilon_{{\rm F}}^{}=\hbar v k_{{\rm F}}^{}$ the Fermi energy), $\partial f_{{\bf k}}^{0}/\partial \epsilon_{{\bf k}}^{}$ is sharply peaked around the Fermi level and the current density can thus be obtained by performing the angular average of over ${\bf k}$ with the scattering time $\tau_{k_{{\rm F}}^{}}^{}$ evaluated at the Fermi level. As a result, ${\bf j}=e^2_{}v^2_{}\nu_{{\rm F}}^{}\tau_{k_{{\rm F}}^{}}^{}{\bf E}/2$, yielding the 
longitudinal resistivity 
\begin{equation}
\label{Rho}
\rho =\frac{\epsilon_{{\rm F}}^{}}{ne^2 v^2 \tau_{k_{{\rm F}}^{}}^{}}
\end{equation}
with $n=\eta k_{{\rm F}}^{2}/4\pi$ the electronic density.

We now focus on the collision integral $C(\varphi_{{\bf k}}^{})$ for the scattering of electrons on in-plane and flexural phonons in graphene. By comparison with Eq.\ (\ref{Relax}) we will thus deduce the respective transport scattering time, leading to the resistivity via Eq.\ (\ref{Rho}).

%leads to the resistivity $\rho =E_{{\rm F}}^{}/ne^2 v^2 \tau_{k_{{\rm F}}^{}}^{}$
%for $T\ll T_{{\rm F}}^{}$ with $T_{{\rm F}}^{}$ the Fermi temperature. Here $E_{{\rm F}}^{}=\hbar v k_{{\rm F}}^{}$ is the Fermi energy, $n=k_{{\rm F}}^{2}/\pi$ is the electronic density and $1/\tau_{k_{{\rm F}}^{}}^{}$ is the transport scattering rate for electrons at the Fermi level.
%Although, strictly speaking, phonon scattering is inelastic, acoustic phonons in graphene have a very low group velocity with respect to the electronic one, which justifies a quasielastic approximation for the scattering rate. 
%The latter is thus given by the standard Fermi golden rule expression
%\begin{equation}\label{Golden}
%\frac{1}{\tau_{k_{{\rm F}}^{}}^{}} =\frac{2\pi}{\hbar}\sum_{f}^{}\left| M_{fi}^{}\right|^{2}_{}\left( 1-\cos\theta\right) \delta (\epsilon_{{\bf k}_{{\rm in}}^{}}^{}-\epsilon_{{\bf k}_{{\rm out}}^{}}^{})
%\end{equation}
%where $M_{fi}^{}$ is the relevant matrix element of the scattering process between the initial and final states $|i\rangle$ and $|f\rangle$ and $\theta$ is the angle between the incoming and outgoing electron momenta. 
%The quasielastic approximation yields the on-shell energy conservation involving only the incoming and outgoing electron energies at the Fermi level. 
%When combined with the single-valley approximation, the quasielasticity implies conservation of chirality. 
%In the following we thus choose $s=1$ without loss of generality.

\subsection{In-plane phonons}
\label{subsec:InPlane}

Let us first consider scattering between electrons and in-plane phonons of type $\nu$ ($\nu=l,\, t$ for longitudinal and transverse modes, respectively). The collision integral $C^{(\nu)}_{}(f_{{\bf k}}^{})$ describing the detailed balance of the occupation of an electronic state with wavevector ${\bf k}$ is given by the Fermi golden rule expression
\begin{widetext}
\begin{eqnarray}
\label{Collision}
&&C^{(\nu)}_{}(f_{{\bf k}}^{})=\frac{2\pi}{\hbar}\sum_{{\bf Q}}^{}W^{(\nu)}_{{\bf k}',{\bf Q};{\bf k}}\left[f_{{\bf k}'}^{}(1-f_{{\bf k}}^{})n_{{\bf Q}}^{(\nu)}-f_{{\bf k}}^{}(1-f_{{\bf k}'}^{})(1+n_{{\bf Q}}^{(\nu)})\right]\delta (\epsilon_{{\bf k}}^{}-\epsilon_{{\bf k}'}^{}-\hbar\omega_{{\bf Q}}^{(\nu)}) +\nonumber \\
&&\quad\quad\quad\quad\; +\frac{2\pi}{\hbar}\sum_{{\bf Q}}^{} W^{(\nu)}_{{\bf k}';{\bf k},{\bf Q}}\left[f_{{\bf k}'}^{}(1-f_{{\bf k}}^{})(1+n_{{\bf Q}}^{(\nu)})-f_{{\bf k}}^{}(1-f_{{\bf k}'}^{})n_{{\bf Q}}^{(\nu)}\right] \delta (\epsilon_{{\bf k}}^{}-\epsilon_{{\bf k}'}^{}+\hbar\omega_{{\bf Q}}^{(\nu)})\; ,
\end{eqnarray}
\end{widetext}
where $n_{{\bf Q}}^{(\nu)}$ is the distribution function for phonons and the factors $W^{(\nu)}_{{\bf k}';{\bf k},{\bf Q}}$ are given by
\begin{equation}
\label{WW}
W^{(\nu)}_{{\bf k}';{\bf k},{\bf Q}}=\delta_{{\bf k}',{\bf k}+{\bf Q}}^{}\big|\langle {\bf k}'|w^{(\nu)}_{{\bf Q}}|{\bf k}\rangle\big|^{2}_{}\; .
\end{equation}
The first term describes the process of absorption of a phonon with wavevector ${\bf Q}$ and energy $\hbar\omega^{(\nu)}_{{\bf Q}}$ by an electron with wavevector ${\bf k}'$ and the reverse process involving the emission of the phonon by an electron with wavevector {\bf k}. In parallel, the second term describes the emission of the phonon by the electron in state $|{\bf k}'\rangle$ and the absorption of the phonon by the electron in $|{\bf k}\rangle$. 
%Momentum conservation yields ${\bf k}={\bf k}'+{\bf Q}$ in the first term and ${\bf k}'={\bf k}+{\bf Q}$ in the second. 

We now proceed to linearize the collision integral above, \cite{Landau10} making use of Eq. (\ref{Varphi}). In doing so, we expand $f_{{\bf k}'}^{0}-f_{{\bf k}}^{0}=\pm \hbar\omega_{{\bf Q}}^{(\nu)}\left(\partial f_{{\bf k}}^{0}/\partial \epsilon_{{\bf k}}^{}\right)$ for $\epsilon_{{\bf k}'}^{}=\epsilon_{{\bf k}'}^{}\pm\hbar\omega_{{\bf Q}}^{(\nu)}$, valid if $\hbar\omega_{{\bf Q}}^{(\nu)}\ll \epsilon_{{\bf k}}^{}$.
Indeed, although, strictly speaking, phonon scattering is inelastic, acoustic phonons in graphene have a low group velocity with respect to the electrons, which justifies a quasielastic approximation for the scattering rate. We also assume the phonons to always remain in equilibrium. \cite{Landau10} 
%(i.e. $\chi_{{\bf Q}}^{(\nu)}=0$). 
Summing up the contributions in Eq.\ (\ref{Collision}) we obtain the linearized collision integral \cite{Landau10}
\begin{eqnarray}
\label{CollLin2}
&&C^{(\nu)}_{}(\varphi_{{\bf k}}^{})\simeq 
\frac{2\pi}{\hbar}\sum_{{\bf Q}}^{}2\omega_{{\bf Q}}^{(\nu)}
\frac{\partial n_{{\bf Q}}^{0(\nu)}}{\partial \omega_{{\bf Q}}^{(\nu)}}
\frac{\partial f_{{\bf k}}^{0}}{\partial \epsilon_{{\bf k}}^{}}
\left(\varphi_{{\bf k}+{\bf Q}}^{}-\varphi_{{\bf k}}^{}\right)\times \nonumber \\
&&\quad\quad\quad\quad\quad\quad\times\big|\langle {\bf k}+{\bf Q}|w^{(\nu)}_{{\bf Q}}|{\bf k}\rangle\big|^{2}_{}
\delta (\epsilon_{{\bf k}+{\bf Q}}^{}-\epsilon_{{\bf k}}^{})\; .
\end{eqnarray}
where $n_{{\bf Q}}^{0(\nu)}=(\exp [\hbar\omega_{{\bf Q}}^{(\nu)}/k_{{\rm B}}^{}T]-1)^{-1}_{}$ is the equilibrium Bose-Einstein distribution.
As the standard solution (\ref{BoltzSol}) yields $\varphi_{{\bf k}}^{}\propto \cos\zeta$, we can write 
\begin{eqnarray}
\label{CollLinFin}
&&C^{(\nu)}_{}(\varphi_{{\bf k}}^{})\simeq -\varphi_{{\bf k}}^{}\frac{\partial f_{{\bf k}}^{0}}{\partial \epsilon_{{\bf k}}^{}}\cdot
\frac{2\pi}{\hbar}\sum_{{\bf Q}}^{}2\omega_{{\bf Q}}^{(\nu)}
\frac{\partial n_{{\bf Q}}^{0(\nu)}}{\partial \omega_{{\bf Q}}^{(\nu)}}
\left(1-\cos\theta\right)\times \nonumber \\
&&\quad\quad\quad\quad\quad\quad\times\big|\langle {\bf k}+{\bf Q}|w^{(\nu)}_{{\bf Q}}|{\bf k}\rangle\big|^{2}_{}
\delta (\epsilon_{{\bf k}+{\bf Q}}^{}-\epsilon_{{\bf k}}^{})\; ,
\end{eqnarray}
with $\theta$ the angle between the electronic wavevectors ${\bf k}$ and ${\bf k}+{\bf Q}$. By direct comparison with Eq.\ (\ref{Relax}), we deduce the transport scattering rate
\begin{eqnarray}\label{Golden}
&&\frac{1}{\tau_{{\bf k}}^{}} =
-\frac{2\pi}{\hbar}\sum_{{\bf Q}}^{}2\omega_{{\bf Q}}^{(\nu)}
\frac{\partial n_{{\bf Q}}^{0(\nu)}}{\partial \omega_{{\bf Q}}^{(\nu)}}
\left(1-\cos\theta\right)\times \nonumber \\
&&\quad\quad\quad\quad\quad\quad\times\big|\langle {\bf k}+{\bf Q}|w^{(\nu)}_{{\bf Q}}|{\bf k}\rangle\big|^{2}_{}
\delta (\epsilon_{{\bf k}+{\bf Q}}^{}-\epsilon_{{\bf k}}^{})\; .
\end{eqnarray}
The derivative of the Bose distribution implies that the relevant phonons to be considered have energies up to $\hbar\omega_{{\bf Q}}^{(\nu)}\sim k_{{\rm B}}^{}T$ and their wavenumbers restricted to $Q\lesssim q_{T}^{(\nu)}$, with $\hbar\omega_{q_T^{(\nu)}}^{(\nu)}=k_{{\rm B}}^{}T$. 
This can be understood as follows. In absorption processes, it is only these phonons that have a large enough equilibrium occupation, while in emission processes, electrons can transfer at most an energy of order $k_{{\rm B}}^{}T$ to the phonons due to the Fermi distribution.

\begin{figure}[ht]
	\centering
		\includegraphics[width=0.5\columnwidth]{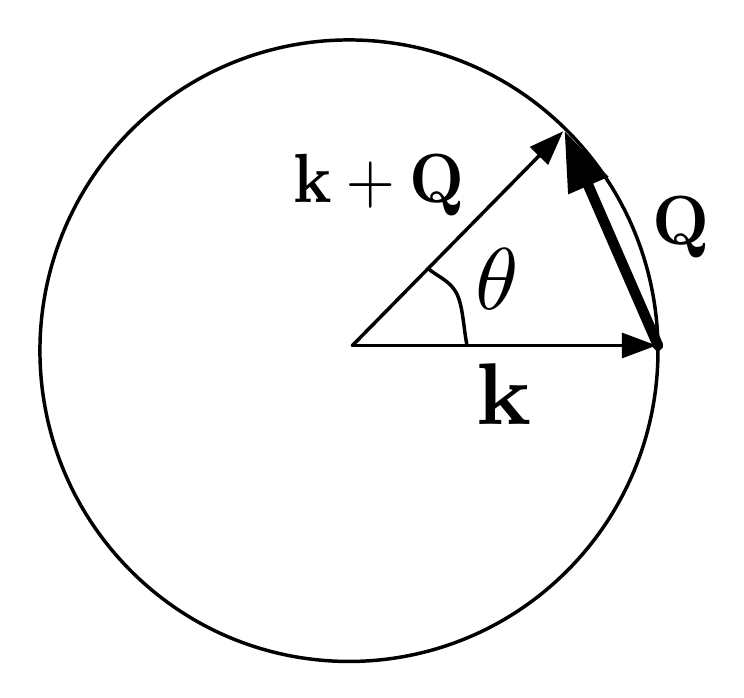}
			\caption{The on-shell condition forces the incoming and outgoing electron momenta ${\bf k}$ and ${\bf k}+{\bf Q}$ on the Fermi circle, with ${\bf Q}$ the phonon-induced scattering wavevector.
			\label{OnShell}}
\end{figure}

In addition, the quasielastic approximation in Eq.\ (\ref{Golden}) demands  energy conservation involving only the incoming and outgoing electron energies at the Fermi level. When combined with the single-valley approximation, the quasielasticity implies conservation of chirality. 
In the following, we thus choose $s=1$ without loss of generality.
The on-shell condition and the relation between momenta and their angles (sketched in Fig.\ \ref{OnShell}) yield 
\begin{equation}\label{Shell}
\delta (\epsilon_{{\bf k}+{\bf Q}}^{}-\epsilon_{{\bf k}}^{})=\sum_{\Phi_0}^{}\frac{1}{\hbar vQ|\cos (\theta /2)|}\, \delta (\Phi -\Phi_0)\; ,
\end{equation}
where $\Phi_0$ denotes the angles between ${\bf k}$ and ${\bf Q}$ fulfilling $\cos \Phi_0 =-\sin (\theta /2)=-Q/2k_{{\rm F}}^{}$. This condition also forces $Q\in [0,2k_{{\rm F}}^{}]$. As a result we deduce 
\begin{equation}\label{Angle}
1-\cos \theta = 2 \sin^{2}_{}(\theta /2)=2\left(\frac{Q}{2k_{{\rm F}}^{}}\right)^{2}_{}\; .
\end{equation}
\\
%can then be calculated as 
%\begin{equation}\label{RateIn}
%\frac{1}{\tau_{k_{{\rm F}}^{}}^{}}=\frac{2\pi}{\hbar}\sum_{{\bf Q}}^{}8g_{1}^{2}Q^2 \xi_{{\bf Q}}^{2}n_{{\bf Q}}^{}\, \frac{\sin^2 (\theta /2)|\cos(\theta /2)|}{\hbar v Q}\, \delta (\Phi -\Phi_0)\; .
%\end{equation}

We thus conclude that the phonon-induced resistivity exhibits a number of temperature regimes in suspended graphene, for two reasons: The first reason is conventional and is associated with the Bloch-Gr\"uneisen temperature $T_{{\rm BG}}^{(\nu)}=\hbar \omega^{(\nu)}_{2k_{{\rm F}}^{}}/k_{{\rm B}}^{}$.
When $T\gg T_{\rm BG}^{(\nu)}$, large angle scattering is possible while, at $T\ll T_{\rm BG}^{(\nu)}$, only small-angle scattering is relevant. The second reason is specific to graphene and is associated with the screening dependence of the relevant electron-phonon coupling mechanism. If $T\gg T_{\rm BG}^{(\nu)}$, screening hardly affects the dominant deformation potential contribution to the resistivity, while, at small enough temperatures, it suppresses the deformation potential in favour of the gauge field coupling.
%We first briefly describe the resulting regimes before we present more detailed results.

%Due to the wavevector dependence of $1-\cos \theta$ and of the screening factor in Eq.\ (\ref{Screening}), the rate (\ref{Golden}) shows various dependences on temperature if small (vs large) angle scattering dominates or if the screened deformation potential is dominated by the gauge field coupling.
%
Indeed, in the high temperature regime $T\gg T_{{\rm BG}}^{(\nu)}$ the sum over phonon momenta is cut off at $2k_{{\rm F}}^{}$ and scattering at all angles is possible. Consequently, screening does not significantly affect the deformation potential coupling. The latter yields the dominant contribution to the resistivity for longitudinal phonons with $g_{1}^{({\rm sc})}(Q)\simeq g_1^{}$ while transverse modes depend only on the unscreened gauge field coupling. 

In contrast, in the regime $T\ll T_{{\rm BG}}^{(\nu)}$, the sum is cut off at $q_T^{(\nu)}$ (with $q_T^{(\nu)}\ll 2k_{{\rm F}}^{}$). This constraint implies dominant small angle scattering ($\theta\ll 1$), and the term $1-\cos\theta$ therefore yields the factor $2(q_T^{(\nu)}/2k_{{\rm F}}^{})^{2}_{}\sim (T/T_{{\rm BG}}^{(\nu)})^{2}_{}\ll 1$.
In addition, in this regime the deformation potential coupling for longitudinal phonons is strongly screened. We thus identify an intermediate regime for $T_{{\rm GD}}^{(l)}<T<T_{{\rm BG}}^{(l)}$ (with $T_{{\rm GD}}^{(l)}=\hbar\omega^{{(l)}}_{Q_{{\rm GD}}^{}}/k_{{\rm B}}^{}$) where small angle scattering is accompanied by a dominant deformation potential. Finally, at low-temperatures $T<T_{{\rm GD}}^{(l)}$ the scattering rate is dominated by the unscreened gauge field coupling. As screening does not affect the coupling for transverse phonons, the latter do not show an intermediate temperature regime.

%This implies that the rate (\ref{Golden}) depends differently on temperature for $T\ll T_{{\rm BG}}^{(\nu)}$ and $T\gg T_{{\rm BG}}^{(\nu)}$, with $T_{{\rm BG}}^{(\nu)}=\hbar \omega^{(\nu)}_{2k_{{\rm F}}^{}}/k_{{\rm B}}^{}$ the Bloch-Gr\"uneisen temperature.
%
%In the low-temperature regime $T\ll T_{{\rm BG}}^{(\nu)}$, the sum over phonon momenta is cut off at $q_T^{(\nu)}$ (with $q_T^{(\nu)}\ll 2k_{{\rm F}}^{}$), yielding dominant small angle scattering ($\theta\ll 1$). The term $1-\cos\theta$ therefore yields the factor $2(q_T^{(\nu)}/2k_{{\rm F}}^{})^{2}_{}\sim (T/T_{{\rm BG}}^{(\nu)})^{2}_{}\ll 1$.
%In addition, in this regime the deformation potential coupling for longitudinal phonons is strongly screened and its contribution to the resistivity is dominated by the unscreened gauge field coupling for $T<T_{{\rm GD}}^{(l)}$, with $T_{{\rm GD}}^{(l)}=\hbar\omega^{{(l)}}_{Q_{{\rm GD}}^{}}/k_{{\rm B}}^{}$.
%
%In contrast, in the high temperature regime $T\gg T_{{\rm BG}}^{(\nu)}$ the sum is cut off at $2k_{{\rm F}}^{}$ and scattering at all angles is possible. Consequently, screening does not significantly affect the deformation potential coupling, which thus yields the dominant contribution to the resistivity for longitudinal phonons with $g_{1}^{({\rm sc})}(Q)\simeq g_1^{}$.

\subsubsection{Longitudinal phonons} 
We first turn to the transport scattering rate due to longitudinal in-plane phonons which couple to the carriers through both, a gauge field and the deformation potential. The deformation potential contribution to the scattering rate has been analysed before for monolayer \cite{Pietronero,Stauber,Hwang08} and bilayer graphene. \cite{Heikkila}
Note however, that while the deformation potential is believed to have a larger bare coupling constant, it is suppressed at long wavelengths due to screening for nonzero doping.  Thus, screening leads to temperature regimes in addition to those associated with the Bloch-Gr\"uneisen temperature.

With the coupling matrix in Eq.\ (\ref{W}) we deduce
\begin{eqnarray}
&&\langle {\bf k}+{\bf Q} |w_{{\bf Q}}^{(l)}| {\bf k}\rangle =\frac{iQ\xi_{{\bf Q}}^{(l)}}{2}\times \nonumber \\
&&\times\left[g_{1}^{({\rm sc})}(Q)\left(1+e^{-i\theta}_{}\right)-ig_{2}^{}\left(e^{i(2\Phi +\phi_{{\bf k}}^{})}_{}-e^{-i(2\Phi +\phi_{{\bf k}}^{}+\theta)}_{}\right)\right]
\nonumber
\end{eqnarray}
including both deformation potential and gauge field couplings.
The dominant contributions to the scattering rate (\ref{Golden}) in the various temperature regimes therefore are
%In these regimes one therefore obtains the Bloch law for in-plane modes \cite{Harrison,Hwang08}
\begin{eqnarray}\label{RateInTLong}
&&\frac{1}{\tau_{k_{{\rm F}}^{}}^{}}\simeq \frac{32\pi^3}{15}\, \Gamma' \left(\frac{T}{T_{{\rm BG}}^{(l)}}\right)^{4}_{}\quad\quad\quad\quad\;\;\; {\rm for}\; T\ll T_{{\rm GD}}^{(l)}\nonumber\\
&&\frac{1}{\tau_{k_{{\rm F}}^{}}^{}}\simeq 12\pi^5 \Gamma \left(\frac{2k_{{\rm F}}^{}}{Q_{{\rm TF}}^{}}\right)^{2}_{}\left(\frac{T}{T_{{\rm BG}}^{(l)}}\right)^{6}_{}\quad\;{\rm for}\; T_{{\rm GD}}^{(l)}\ll T\ll T_{{\rm BG}}^{(l)}\nonumber\\
&&\frac{1}{\tau_{k_{{\rm F}}^{}}^{}}\simeq \Gamma f(\frac{Q_{{\rm TF}}^{}}{2k_{{\rm F}}^{}})\frac{T}{T_{{\rm BG}}^{(l)}}\quad\quad\quad\quad\quad\quad\;\, {\rm for}\; T\gg T_{{\rm BG}}^{(l)}
\end{eqnarray}
with $\Gamma = g_1^2k_{{\rm F}}^{2}/2\hbar\rho_0 vv^{(l)}_{}$, $\Gamma'=\Gamma g^{2}_{2}/g_{1}^{2}$ and $f(x)$ a function fulfilling $f(x\ll 1)\simeq 1-32\, x/3\pi$ and $f(x\gg 1)\simeq 1/2x^{2}_{}$. %Similar results are obtained for the phonon-emission case, as the relevant phonon energies are bound by temperature by the fermionic phase space argument.
 
The $T^4$ dependence for $T\ll T_{{\rm GD}}^{(l)}$ is due to the gauge field coupling. It is analogous to the $T^5_{}$ dependence (Bloch law) for $T\ll T_{{\rm BG}}^{(l)}$ in three dimensions, \cite{Harrison} the difference in the power law stemming from the reduced dimensionality of momentum space.\\
In the intermediate temperature regime, $T_{{\rm GD}}^{(l)}\ll T\ll T_{{\rm BG}}^{(l)}$, the screened deformation potential dominates over the gauge field coupling. The $T^6_{}$ dependence originates from the combined effects of screening and small angle scattering.\\
Finally, the linear temperature dependence for $T\gg T_{{\rm BG}}^{(l)}$ stems from the high-temperature expansion of the Bose distribution.
Since the typical momentum transfers in this regime are of order $2k_{{\rm F}}^{}$, it is associated with the essentially unscreened deformation potential.
This regime turns out to be the most relevant for the interpretation of recent measurements. \cite{Fuhrer,Bolotin}
%
%yields
%\begin{eqnarray}\label{RateInTGauge}
%&&\frac{1}{\tau_{k_{{\rm F}}^{}}^{}}\Big|_{{\rm Gauge}}^{}\simeq \frac{2\pi^3}{15} \Gamma_{{\rm G}}^{} \left(\frac{T}{T_{{\rm BG}}^{(l)}}\right)^{4}_{}\quad\quad{\rm for}\; T\ll T_{{\rm BG}}^{(l)}\nonumber\\
%&&\frac{1}{\tau_{k_{{\rm F}}^{}}^{}}\Big|_{{\rm Gauge}}^{}\simeq \Gamma_{{\rm G}}^{} \frac{T}{T_{{\rm BG}}^{(l)}}\quad\quad\quad\quad\; {\rm for}\; T\gg T_{{\rm BG}}^{(l)}
%\end{eqnarray}
%with $\Gamma_{{\rm G}}^{} = g_2^2 k_{{\rm F}}^{2}/\hbar\rho_0 vv^{(l)}_{}$. The $T^4$ dependence at $T\ll T_{{\rm BG}}^{(l)}$ differs from the analogous $T^5_{}$ law in 3D \cite{Harrison} due to the reduced dimensionality of the momentum space.
%
%Thus the transport scattering rate due to scattering of electrons with longitudinal phonons shows three limiting behaviours. For $T\ll T_{{\rm GD}}^{(l)}$ it depends on temperature as $T^4$ and is mostly due to the gauge-field coupling. For $T_{{\rm GD}}^{(l)}\ll T\ll T^{(l)}_{{\rm BG}}$ the rate is dominated by the screened deformation potential and shows a $T^6$ dependence. Finally, for $T\gg T^{(l)}_{{\rm BG}}$ it is dominated by the essentially unscreened deformation potential, with a linear dependence on temperature. 

\subsubsection{Transverse phonons} 
The analysis above can be extended to transverse in-plane phonons. These have a slightly smaller group velocity than longitudinal modes and couple only via the gauge field mechanism, so that screening is not relevant. As a result, the intermediate regime in Eq.\ (\ref{RateInTLong}) is absent. The coupling term in Eq. (\ref{W}) yields
\begin{equation}
\langle {\bf k}+{\bf Q} |w_{{\bf Q}}^{(t)}| {\bf k}\rangle =\frac{ig_{2}^{}Q\xi_{{\bf Q}}^{(t)}}{2}\left(e^{i(2\Phi +\phi_{{\bf k}}^{})}_{}+e^{-i(2\Phi +\phi_{{\bf k}}^{}+\theta)}_{}\right)\; ,
\nonumber
\end{equation}
leading to the scattering rate
\begin{eqnarray}\label{RateInTTrans}
&&\frac{1}{\tau_{k_{{\rm F}}^{}}^{}}\simeq \frac{16\pi^3}{15} \Gamma_{{\rm G}}^{} \left(\frac{T}{T_{{\rm BG}}^{(t)}}\right)^{4}_{}\quad\quad{\rm for}\; T\ll T_{{\rm BG}}^{(t)}\nonumber\\
&&\frac{1}{\tau_{k_{{\rm F}}^{}}^{}}\simeq \Gamma_{{\rm G}}^{} \frac{T}{T_{{\rm BG}}^{(t)}}\quad\quad\quad\quad\quad\quad\;\; {\rm for}\; T\gg T_{{\rm BG}}^{(t)}
\end{eqnarray}
with $\Gamma_{{\rm G}}^{} = g_2^2 k_{{\rm F}}^{2}/\hbar\rho_0 vv^{(t)}_{}$.
The transverse in-plane phonons thus yield a contribution to the resistivity comparable to the longitudinal ones for $T\ll T_{{\rm GD}}^{(l)}$, while for higher temperatures they can be neglected to a good approximation.

\subsubsection{Temperature-dependent resistivity in recent experiments}  
Recent experiments in graphene highlighted the contribution to the resistivity due to electron-phonon scattering $\Delta\rho$ at $T\gg T_{{\rm BG}}^{(l)}$. \cite{Fuhrer,Bolotin} In this  regime, using our estimates above, the resistivity due to scattering by in-plane phonons, $\Delta\rho_{{\rm in}}^{} =\epsilon_{{\rm F}}^{}/ne^2 v^2 \tau_{k_{{\rm F}}^{}}^{}$, is dominated by the deformation potential coupling of longitudinal modes and is given by
\begin{equation}\label{RhoIn}
\Delta\rho_{{\rm in}}^{}\simeq \frac{\pi g_{1}^{2}f(Q_{{\rm TF}}^{}/2k_{{\rm F}}^{})}{4\hbar \rho_{0}^{}e^2 v^2 v^{(l)2}_{}}\, k_{{\rm B}}^{}T\simeq \frac{h}{e^{2}_{}}\, 2\cdot 10^{-8}_{}\, \tilde{g}_{1}^{2}\, \tilde{T}
\end{equation}
in terms of the rescaled quantities $\tilde{g}_{1}^{}=g_{1}^{}f(Q_{{\rm TF}}^{}/2k_{{\rm F}}^{})/{\rm eV}$ and $\tilde{T}=T/{\rm K}$.
The contribution to the resistivity due to in-plane phonons is thus linear in temperature and independent of the electron concentration for $T\gg T_{{\rm BG}}^{(l)}$. 

Experiments in non-suspended samples \cite{Fuhrer} show a temperature-dependent component of the resistivity compatible with Eq.\ (\ref{RhoIn}), which is quantitatively consistent with the estimates for the bare deformation potential coupling constant. \cite{Suzuura}
Suspended graphene devices \cite{Bolotin} likewise show a linear $T$ dependence, compatible with in-plane phonons, but accompanied by an unexpected electron-density dependence not captured by our analysis above. In particular, for increasing electron concentration the resistivity has been observed to decrease and finally saturate to the in-plane phonon contribution (\ref{RhoIn}) at large densities. 

While it may be tempting to think that this behaviour originates from the screening of the deformation potential at increasing electron density, our analysis shows that this is not the case.  
%\cite{FelixPacoEros}. We briefly discuss this issue before turning to the analysis of flexural modes.
%
%{\em Screening} --- The deformation potential, as any scalar potential in the Dirac hamiltonian, is affected by screening. In the long-wavelength limit the corresponding coupling constant $g_{1}^{}$ is then reduced by the screening factor
%\begin{equation}\label{Screening}
%\frac{1}{1+{\cal V}_{{\bf Q}}^{}\Pi_{{\bf Q}}^{}}\simeq \frac{Q}{Q+Q_{{\rm TF}}^{}}
%\end{equation}
%with ${\cal V}_{{\bf Q}}^{}$ the 2D Coulomb interaction, $\Pi_{{\bf Q}}^{}$ the fermionic polarization, $Q_{{\rm TF}}^{}=2\pi e^2 \nu$ the Thomas-Fermi screening wavevector and $\nu\propto k_{{\rm F}}^{}$ the electronic DOS in graphene. 
Indeed, for $T>T_{{\rm BG}}^{(l)}$, $Q$ is cut off at $2k_{{\rm F}}^{}$. This, together with the fact that $Q_{{\rm TF}}^{}\propto k_{{\rm F}}^{}$, implies that the screening factor in (\ref{Screening}) does not introduce any additional density-dependence into the scattering rate. 
%Thus screening of the deformation potential coupling cannot be the reason for the observed electron-density dependence in the resistivity of suspended samples. 
Moreover, if this was a relevant effect, it should appear in non-suspended samples as well.

%Screening can have a non-trivial effect in the low temperature regime $q_{T}^{}\ll Q_{{\rm TF}}^{}$. In this case, momenta are bound by $q_{T}^{}$ and the screening factor (\ref{Screening}) significantly reduces the contribution to the resistivity due to the deformation potential. Since, in parallel, the gauge field coupling is not affected by screening, we predict that, in the $T<T_{{\rm BG}}^{({\rm in})}$ a further crossover temperature $T_{{\rm G-D}}^{}$ emerges, below which the resistivity is mostly due to gauge-field coupling and above which it is dominated by the deformation potential.
%
%Our analysis shows that the in-plane phonon contribution to the resistivity at high temperatures is independent of density even in presence of screening. 
The in-plane phonon contribution (\ref{RhoIn}) seems sufficient to describe the behaviour of non-suspended samples, where flexural (out-of-plane) deformations are suppressed by the direct contact with a substrate.
In graphene samples on very rough substrates small out-of-plane fluctuations could still survive due to the non-perfect adhesion of the membrane to the surface of the substrate.

In contrast, flexural phonons can become relevant in suspended devices. 
%Indeed, their contribution to the resistivity has been shown to be dominant with respect to the in-plane one at low temperatures \cite{Mariani}. 
We now turn to analyse their effect along the same line as for the in-plane modes above. 
%The discussion of the transport properties related to flexural modes may shed light on the qualitative difference between the density dependence observed in suspended vs non-suspended samples.
 
\subsection{Flexural phonons}
\label{subsec:Flexural}

The analysis of the resistivity due to flexural phonon modes is more involved due to the presence of two phonons with wavevectors ${\bf q}_1^{}$ and ${\bf q}_2^{}$ at each interaction vertex. These yield four possible processes involving double absorption, double emission or mixed absorption-emission terms in the scattering rate. In addition, four relevant wavevector scales come into play, namely $q_T^{(h)}$, $Q_{{\rm GD}}^{}$, $q_*^{}$ and $k_{{\rm F}}^{}$, related to temperature ($\hbar\omega_{q_T^{(h)}}^{(h)}=k_{{\rm B}}^{}T$), screening, tension, and electron concentration, respectively. A further ultraviolet scale, corresponding to a Debye energy of order $\hbar\omega^{(h)}_{1/a}\sim 100\, {\rm meV}$, is assumed to provide the largest energy scale. For temperatures higher than $\hbar\omega^{(h)}_{1/a}/k_{{\rm B}}^{}$ the straightforward expansion of the two Bose distributions yields a scattering rate depending on temperature as $T^2$. However, at those energies the elastic treatment of the membrane as well as the elastic approximation in the scattering rate may be questionable. We thus focus on lower temperatures which are relevant to experiments.

The collision integral including all emission and absorption processes for scattering of electrons and flexural phonons has the form
\begin{widetext}
\begin{eqnarray}
\label{CollisionFlex}
&&C^{(h)}_{}(f_{{\bf k}}^{})=\frac{2\pi}{\hbar}\sum_{{\bf q}_{1}^{},{\bf q}_{2}^{}}^{}W^{(h)}_{{\bf k};{\bf k}',{\bf q}_{1}^{},{\bf q}_{2}^{}} \left[f_{{\bf k}'}^{}(1-f_{{\bf k}}^{})n_{{\bf q}_{1}^{}}^{(h)}n_{{\bf q}_{2}^{}}^{(h)}
-f_{{\bf k}}^{}(1-f_{{\bf k}'}^{})(1+n_{{\bf q}_{1}^{}}^{(h)})(1+n_{{\bf q}_{2}^{}}^{(h)})\right]\, \delta (\epsilon_{{\bf k}}^{}-\epsilon_{{\bf k}'}^{}-\hbar\omega_{{\bf q}_{1}^{}}^{(h)}-\hbar\omega_{{\bf q}_{2}^{}}^{(h)})+\nonumber \\
&&\quad\quad+\frac{2\pi}{\hbar}\sum_{{\bf q}_{1}^{},{\bf q}_{2}^{}}^{}W^{(h)}_{{\bf k},{\bf q}_{1}^{},{\bf q}_{2}^{};{\bf k}'} \left[f_{{\bf k}'}^{}(1-f_{{\bf k}}^{})(1+n_{{\bf q}_{1}^{}}^{(h)})(1+n_{{\bf q}_{2}^{}}^{(h)})
-f_{{\bf k}}^{}(1-f_{{\bf k}'}^{})n_{{\bf q}_{1}^{}}^{(h)}n_{{\bf q}_{2}^{}}^{(h)}\right]\, \delta (\epsilon_{{\bf k}}^{}-\epsilon_{{\bf k}'}^{}+\hbar\omega_{{\bf q}_{1}^{}}^{(h)}+\hbar\omega_{{\bf q}_{2}^{}}^{(h)}) +\nonumber \\
&&\quad\quad+\frac{2\pi}{\hbar}\sum_{{\bf q}_{1}^{},{\bf q}_{2}^{}}^{}W^{(h)}_{{\bf k},{\bf q}_{2}^{};{\bf k}',{\bf q}_{1}^{}} \left[f_{{\bf k}'}^{}(1-f_{{\bf k}}^{})n_{{\bf q}_{1}^{}}^{(h)}(1+n_{{\bf q}_{2}^{}}^{(h)})
-f_{{\bf k}}^{}(1-f_{{\bf k}'}^{})(1+n_{{\bf q}_{1}^{}}^{(h)})n_{{\bf q}_{2}^{}}^{(h)}\right]\, \delta (\epsilon_{{\bf k}}^{}-\epsilon_{{\bf k}'}^{}-\hbar\omega_{{\bf q}_{1}^{}}^{(h)}+\hbar\omega_{{\bf q}_{2}^{}}^{(h)})+\nonumber \\
&&\quad\quad+\frac{2\pi}{\hbar}\sum_{{\bf q}_{1}^{},{\bf q}_{2}^{}}^{}W^{(h)}_{{\bf k},{\bf q}_{1}^{};{\bf k}',{\bf q}_{2}^{}} \left[f_{{\bf k}'}^{}(1-f_{{\bf k}}^{})(1+n_{{\bf q}_{1}^{}}^{(h)})n_{{\bf q}_{2}^{}}^{(h)}
-f_{{\bf k}}^{}(1-f_{{\bf k}'}^{})n_{{\bf q}_{1}^{}}^{(h)}(1+n_{{\bf q}_{2}^{}}^{(h)})\right]\, \delta (\epsilon_{{\bf k}}^{}-\epsilon_{{\bf k}'}^{}+\hbar\omega_{{\bf q}_{1}^{}}^{(h)}-\hbar\omega_{{\bf q}_{2}^{}}^{(h)})
\end{eqnarray}
\end{widetext}
with
\begin{equation}
\label{WWh}
W^{(h)}_{{\bf k};{\bf k}',{\bf q}_{1}^{},{\bf q}_{2}^{}}=\delta_{{\bf k},{\bf k}'+{\bf q}_{1}^{}+{\bf q}_{2}^{}}^{}\big|\langle {\bf k}|w^{(h)}_{{\bf q}_{1}^{},{\bf q}_{2}^{}}|{\bf k}'\rangle\big|^{2}_{}\; .
\end{equation}
A lengthy but straightforward calculation analogous to the case of in-plane modes above yields the linearized collision integral of the form (\ref{Relax}), with the transport scattering rate
\begin{eqnarray}
\label{GoldenH}
&&\frac{1}{\tau_{{\bf k}}^{}}= 
\frac{2\pi}{\hbar}\sum_{{\bf q}_{1}^{},{\bf q}_{2}^{}}^{}2
\frac{\partial n_{{\bf q}_{1}^{}}^{0(h)}}{\partial \omega_{{\bf q}_{1}^{}}^{(h)}}
\frac{\partial n_{{\bf q}_{2}^{}}^{0(h)}}{\partial \omega_{{\bf q}_{2}^{}}^{(h)}}\,\frac{k_{{\rm B}}^{}T}{\hbar}\,
\left(1-\cos\theta\right)\times \nonumber \\
&&\quad\quad\quad\times\big|\langle {\bf k}+{\bf Q}|w^{(h)}_{{\bf q}_{1}^{},{\bf q}_{2}^{}}|{\bf k}\rangle\big|^{2}_{}
\delta (\epsilon_{{\bf k}+{\bf Q}}^{}-\epsilon_{{\bf k}}^{})\times \\
&&\quad\quad\times\left[\frac{\omega_{{\bf q}_{1}^{}}^{(h)}+\omega_{{\bf q}_{2}^{}}^{(h)}}{1+n_{{\bf q}_{1}^{}}^{0(h)}+n_{{\bf q}_{2}^{}}^{0(h)}}+\frac{\omega_{{\bf q}_{1}^{}}^{(h)}-\omega_{{\bf q}_{2}^{}}^{(h)}}{n_{{\bf q}_{2}^{}}^{0(h)}-n_{{\bf q}_{1}^{}}^{0(h)}}\right]\; ,\nonumber
\end{eqnarray}
with $\theta$ the angle between the electronic wavevectors ${\bf k}$ and ${\bf k}+{\bf Q}$ and ${\bf Q}={\bf q}_{1}^{}+{\bf q}_{2}^{}$ the total wavevector transferred to the phonons in the scattering process.

The general considerations in Sec.\ \ref{subsec:InPlane} hold for flexural phonons as well. In particular, the on-shell condition 
(\ref{Shell}) as well as Eq.\ (\ref{Angle}) remain valid. The matrix element of the electron-phonon coupling is given from Eq.\ (\ref{W}) as
\begin{eqnarray}\label{CoupFlex}
&&\langle {\bf k} |w_{{\bf q}_{1}^{},{\bf q}_{2}^{}}^{(h)}| {\bf k}+{\bf Q}\rangle =-\frac{q_1^{}q_2^{}\xi_{{\bf q}_{1}^{}}^{(h)}\xi_{{\bf q}_{2}^{}}^{(h)}}{4} \Big[ g_{1}^{({\rm sc})}(Q)\cos \phi\, \left(1+e^{i\theta}_{}\right) \nonumber \\
&&\quad\quad\quad\quad\quad\quad\quad\quad\quad\quad -ig_{2}^{}\left(e^{i(\psi+\phi_{{\bf k}}^{}+\theta)}_{}-e^{-i(\psi+\phi_{{\bf k}}^{})}_{}\right)\Big]\; .
\nonumber
\end{eqnarray}
%In analogy to what shown above, the two-phonon scattering rate due to all four absorption/emission processes is given by
%\begin{eqnarray}\label{RateFlex}
%&&\frac{1}{\tau_{k_{{\rm F}}^{}}^{}}=\frac{2\pi}{\hbar}\sum_{{\bf q}_1,{\bf q}_2}^{}4g_{1}^{2} q_1^2 q_2^2\xi_{{\bf q}_1}^{2} \xi_{{\bf q}_2}^{2}n_{{\bf q}_1}^{}n_{{\bf q}_2}^{}\times\nonumber \\
%&&\quad\quad\quad\times\frac{\sin^2 (\theta/2)|\cos (\theta/2)|}{\hbar v |{\bf Q}|}\, \delta (\Phi -\Phi_0)\; 
%\end{eqnarray}
%with $\Phi$ the angle between ${\bf Q}={\bf q}_1+{\bf q}_2$ and ${\bf k}$, and $\cos \Phi_0 =-\sin (\theta /2)=-Q/2k_{{\rm F}}^{}$. As in the case above, here we considered only the dominant contribution due to thermal phonons and the geometrical interpretation of the angle $\theta$ is as in Figure \ref{OnShell}.

Temperature enters the scattering rate (\ref{Golden}) via the Bose distributions and cuts off the relevant phonon momenta at $q_1 ,\, q_2\lesssim q_T^{(h)}$. Contrary to what happens for in-plane modes, this is true even for $q_T^{(h)}\gg 2k_{{\rm F}}^{}$ (i.e. for $T\gg T_{{\rm BG}}^{(h)}$, with $T_{{\rm BG}}^{(h)}=\hbar\omega_{2k_{{\rm F}}^{}}^{(h)}/k_{{\rm B}}^{}$ the Bloch-Gr\"uneisen temperature of flexural phonons). Indeed, while the total scattering wavevector ${\bf Q}$ is bounded by the on-shell condition to fulfil $Q\le 2k_{{\rm F}}^{}$, each individual $q_1$ and $q_2$ can be large, provided that $|{\bf q}_1-{\bf q}_2|\lesssim q_T^{(h)}$. In this regime, scattering at all angles is possible and the dominant electron-phonon coupling mechanism is via the unscreened deformation potential. 

In the opposite limit of low temperature where $T\ll T_{{\rm BG}}^{(h)}$ (i.e. for $q_T^{(h)}\ll 2k_{{\rm F}}^{}$) the scattering rate is dominated by small angle scattering, with  $1-\cos\theta\sim (q_T^{(h)}/2k_{{\rm F}}^{})^{2}_{}\ll 1$. In this regime, screening suppresses the deformation potential coupling in Eq.\ (\ref{CoupFlex}), yielding $g_{1}^{({\rm sc})}(Q)\sim g_{1}^{}q_T^{(h)}/Q_{{\rm TF}}^{}\ll g_{1}^{}$. This suppression favours the unscreened gauge field coupling $g_{2}^{}$ for $q_T^{(h)}\ll Q_{{\rm GD}}^{}$ (i.e. for $T\ll T_{{\rm GD}}^{(h)}$, with $T_{{\rm GD}}^{(h)}=\hbar\omega_{Q_{{\rm GD}}^{}}^{(h)}/k_{{\rm B}}^{}$). 

Tension affects the phonon spectra yielding a linear dispersion for $q<q_*$ and a quadratic one for $q>q_*$. 
%thereby suppressing the phononic DOS and shifting the dominant momenta to the region $q>q_*$.
As the density of phonon states is much higher in the region of the quadratic spectrum, the dispersion relation which dominantly contributes to the resistivity is linear if $q_T^{(h)}\ll q_*$ and quadratic for $q_T^{(h)}\gg q_*$. This DOS suppression at low energy is not accompanied by a change in the flexural-phonon coupling which remains quadratic even in presence of tension as the reflection symmetry with respect to the plane is preserved.  
This is the basic mechanism suppressing the contribution to the resistivity of flexural-phonons with respect to in-plane ones. 
%which then appear as the only relevant modes needed to address the main features in the experimental data. 

It is possible to deduce the relevant temperature and electron concentration dependence of the scattering rate by simple power counting, once the relevant momenta are identified. Indeed, for $q_{}^{}\lesssim q_T^{(h)}$, one has $n_{{\bf q}}^{(h)}\sim k_{{\rm B}}^{}T/\hbar\omega_{{\bf q}}^{(h)}$ and the scattering rate is given by
\begin{eqnarray}
\label{GoldenHAppr}
&&\frac{1}{\tau_{{\bf k}}^{}}\simeq
\frac{2\pi}{\hbar}\sum_{{\bf q}_{1}^{},{\bf q}_{2}^{}}^{}4\frac{k_{{\rm B}}^{2}T^{2}_{}}{ \hbar^{2}_{} \omega_{{\bf q}_{1}^{}}^{(h)}\omega_{{\bf q}_{2}^{}}^{(h)}}
\left(1-\cos\theta\right)\times \\
&&\quad\quad\quad\quad\quad\quad\times\big|\langle {\bf k}+{\bf Q}|w^{(h)}_{{\bf q}_{1}^{},{\bf q}_{2}^{}}|{\bf k}\rangle\big|^{2}_{}
\delta (\epsilon_{{\bf k}+{\bf Q}}^{}-\epsilon_{{\bf k}}^{}) ,\nonumber
\end{eqnarray}
where the factor $4$ accounts for all emission-absorption processes of the two phonons. 
Introducing the total and relative momenta ${\bf Q}$ and ${\bf q}={\bf q}_1-{\bf q}_2$, the transport scattering rate for electrons at the Fermi level scales as 
\begin{eqnarray}\label{RateScaling}
&&\frac{1}{\tau_{k_{{\rm F}}^{}}^{}}\sim T^2\int^{q_{T}^{(h)}} dq\, q\int^{{\rm min}[q_{T}^{(h)},2k_{{\rm F}}^{}]}dQ \times\\
&&\quad\quad\quad\quad\quad\times\left(\frac{Q}{2k_{{\rm F}}^{}} \right)^2\left(\frac{Q}{Q+Q_{{\rm TF}}^{}}\right)^{2\lambda}_{}\frac{q_1^2 q_2^2}{\omega_{{\bf q}_1}^{(h)2}\omega_{{\bf q}_2}^{(h)2}}\nonumber \; .
\end{eqnarray}
Here $\lambda =0$ for $T\ll T_{{\rm GD}}^{(h)}$ when the gauge field coupling is dominant, while $\lambda =1$ for $T\gg T_{{\rm GD}}^{(h)}$ when the screened deformation potential dominates.
In principle seven different regimes can be identified, according to the relative magnitude of the relevant wavevectors.

At low temperatures $T\ll T_{{\rm GD}}^{(h)}\ll T_{{\rm BG}}^{(h)}$, corresponding to $q_T^{(h)}\ll Q_{{\rm GD}}^{}\ll 2k_{{\rm F}}^{}$ , $\lambda =0$ and small angle scattering dominates the scattering rate. We can rescale all wavenumbers by $q_T^{(h)}$ and find two regimes:
\begin{itemize}
\item I: For $q_*^{}\ll q_T^{(h)}$ the relevant dispersion is $\omega\sim q^2$, yielding $q_T^{(h)}\sim \sqrt{T}$ and $1/\tau_{k_{{\rm F}}^{}}^{}\sim T^{5/2}_{}/k_{{\rm F}}^{2}$.
\item II: For $q_T^{(h)}\ll q_*^{}$ the relevant dispersion is $\omega\sim q$, yielding $q_T^{(h)}\sim T$ and $1/\tau_{k_{{\rm F}}^{}}^{}\sim T^{7}_{}/k_{{\rm F}}^{2}$.
\end{itemize}
As pointed out before, \cite{Mariani} the first regime with low tension shows a logarithmic infrared singularity. For a sufficiently large membrane, these singularities are regularised by a temperature-dependent infrared cutoff related to the non-linear coupling of bending and stretching energies in the elastic Lagrangian. \cite{Nelson,Mariani} Small tension or finite size effects yield alternative infrared cutoffs modifying the value of the scattering rates by numerical prefactors of order one.

The expansions above show that, in the first case with very weak tension ($q_*^{}\ll q_T^{(h)}\ll Q_{{\rm GD}}^{}$), the scattering rate due to flexural phonons has a $T^{5/2}_{}$ temperature dependence and is larger than the corresponding one due to in-plane modes, which scales as $T^4_{}$. In contrast, for $q_T^{(h)}\ll q_*^{}\ll Q_{{\rm GD}}^{}$, tension suppresses the flexural contribution to the subdominant $T^7_{}$ term, which allows in-plane scattering to dominate the electron-phonon scattering contribution to the resistivity at low temperatures. 

At intermediate temperatures $T_{{\rm GD}}^{(h)}\ll T\ll T_{{\rm BG}}^{(h)}$, corresponding to $Q_{{\rm GD}}^{}\ll q_T^{(h)}\ll 2k_{{\rm F}}^{}$, small angle scattering still dominates the scattering rate while $\lambda =1$. Screening thus yields an additional factor $(q_T^{(h)}/Q_{{\rm TF}}^{})^{2}_{}$ in the scattering rate.
Rescaling all wavenumbers by $q_T^{(h)}$ we find two regimes similar to the previous analysis:
\begin{itemize}
\item III: For $q_*^{}\ll q_T^{(h)}$ the relevant dispersion is $\omega\sim q^2$, yielding $q_T^{(h)}\sim \sqrt{T}$ and $1/\tau_{k_{{\rm F}}^{}}^{}\sim T^{7/2}_{}/k_{{\rm F}}^{4}$.
\item IV: For $q_T^{(h)}\ll q_*^{}$ the relevant dispersion is $\omega\sim q$, yielding $q_T^{(h)}\sim T$ and $1/\tau_{k_{{\rm F}}^{}}^{}\sim T^{9}_{}/k_{{\rm F}}^{4}$.
\end{itemize}

Finally, in the high temperature limit $T\gg T_{{\rm BG}}^{(h)}$,  corresponding to $q_T^{(h)}\gg 2k_{{\rm F}}^{}$, large angle scattering is possible for $q_1,\, q_2>k_{{\rm F}}^{}$, which yield the dominant contribution to the scattering rate. 
In this regime $\lambda =1$ and the total wavevector $Q$ is cut off at  $2k_{{\rm F}}^{}$ due to the on-shell condition while $q$ is limited by $q_T^{(h)}$. In the case of strong tension $q_{*}^{}\gg q_T^{(h)}$ the suppression of the phononic DOS at low energy shifts the dominant momenta to large values of order $q\sim q_T^{(h)}$. In contrast, small tension $q_{*}^{}\ll q_T^{(h)}$ tends to favour low momenta down to  $q_*^{}$ or $k_{{\rm F}}^{}$, whichever is smaller. We identify three regimes:
\begin{itemize}
\item V: For $q_*^{}\ll 2k_{{\rm F}}^{}\ll q_T^{(h)}$ the relevant dispersion is $\omega\sim q^2$, yielding a dominant contribution for small wavenumbers down to $q\sim k_{{\rm F}}^{}$ resulting in $1/\tau_{k_{{\rm F}}^{}}^{}\sim T^{2}_{}/k_{{\rm F}}^{}$.
\item VI: For $2k_{{\rm F}}^{}\ll q_*^{}\ll q_T^{(h)}$ the relevant dispersion is $\omega\sim q^2$, yielding a dominant contribution for small wavenumbers down to $q\sim q_{*}^{}$ resulting in $1/\tau_{k_{{\rm F}}^{}}^{}\sim T^{2}_{}k_{{\rm F}}^{}$.
\item VII: For $2k_{{\rm F}}^{}\ll q_T^{(h)}\ll q_*^{}$ the relevant dispersion is $\omega\sim q$, yielding $q_T^{(h)}\sim T$ and $1/\tau_{k_{{\rm F}}^{}}^{}\sim T^{4}_{}k_{{\rm F}}^{}$.
\end{itemize}
It has to be noted that the $T^2_{}$ scaling in regime V stems from the relevant $q^2$ dispersion of flexural modes with low tension, and is not trivially obtained from the high-temperature expansion of the two Bose distributions, unlike in the case of in-plane phonons. In this case, in fact, the $q_1$ and $q_2$ momenta are still limited by $q_{T}^{(h)}$.

\begin{figure}[t]
	\centering
		\includegraphics[width=0.9\columnwidth]{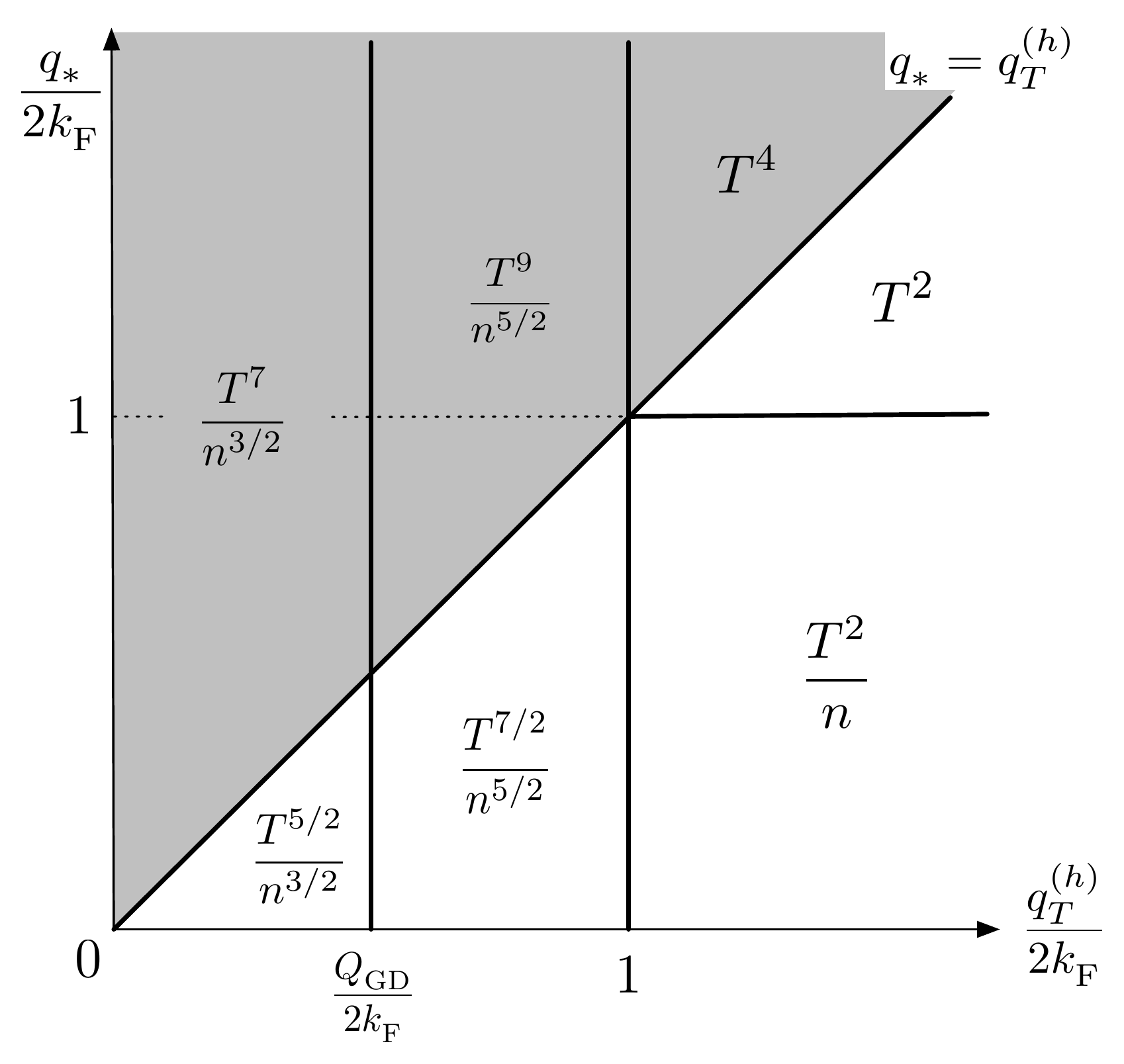}
			\caption{The dependence of the resistivity due to scattering off flexural modes on temperature $T$ and electron density $n$. The grey area identifies the region $q_{*}^{}>q_T^{(h)}$ where the relevant flexural phonon dispersion is dominated by tension and $\omega_{{\bf q}}^{(h)}\simeq \alpha q$. 
			\label{Fig2}}
\end{figure}

The corresponding dependence of the resistivity on temperature and electron density in the seven regions above is summarised in the diagram of Fig.\ \ref{Fig2}.
\\
By a more explicit evaluation of Eq.\ (\ref{GoldenH}) we calculate the scattering rate due to electron-phonon coupling for flexural-modes in the seven regimes above, yielding
\begin{eqnarray}
&&\frac{1}{\tau_{k_{{\rm F}}^{}}^{}}\simeq C_{{\rm G}}^{}\frac{\left (k_{{\rm B}}^{}T\right)^{5/2}_{}}{(\hbar\beta)^{1/2}_{}\beta^{4}_{}(2k_{{\rm F}}^{})^{2}_{}} \quad\quad\quad\; {\rm in\; region\; I}\nonumber\\
&&\frac{1}{\tau_{k_{{\rm F}}^{}}^{}}\simeq C_{{\rm G}}^{}\frac{\left (k_{{\rm B}}^{}T\right)^{7}_{}}{(\hbar\alpha)^{5}_{}\alpha^{4}_{}(2k_{{\rm F}}^{})^{2}_{}} \quad\quad\quad\quad {\rm in\; region\; II} \nonumber\\
&&\frac{1}{\tau_{k_{{\rm F}}^{}}^{}}\simeq C\frac{\left (k_{{\rm B}}^{}T\right)^{7/2}_{}}{(\hbar\beta)^{3/2}_{}\beta^{4}_{}(2k_{{\rm F}}^{}Q_{{\rm TF}}^{})^{2}_{}} \quad\;\;\; {\rm in\; region\; III} \nonumber\\
&&\frac{1}{\tau_{k_{{\rm F}}^{}}^{}}\simeq C\frac{\left (k_{{\rm B}}^{}T\right)^{9}_{}}{(\hbar\alpha)^{7}_{}\alpha^{4}_{}(2k_{{\rm F}}^{}Q_{{\rm TF}}^{})^{2}_{}} \quad\quad\;\; {\rm in\; region\; IV}  \nonumber\\
&&\frac{1}{\tau_{k_{{\rm F}}^{}}^{}}\simeq C\frac{\left (k_{{\rm B}}^{}T\right)^{2}_{}}{\beta^{4}_{}\, 2k_{{\rm F}}^{}} \quad\quad\quad\quad\quad\quad\;\;\;\;\;\; {\rm in\; region\; V}  \nonumber\\
&&\frac{1}{\tau_{k_{{\rm F}}^{}}^{}}\simeq C\frac{\left (k_{{\rm B}}^{}T\right)^{2}_{} 2k_{{\rm F}}^{}}{\beta^{4}_{}q^{2}_{*}} \quad\quad\quad\quad\quad\;\;\; {\rm in\; region\; VI}  \nonumber\\
&&\frac{1}{\tau_{k_{{\rm F}}^{}}^{}}\simeq C\frac{\left (k_{{\rm B}}^{}T\right)^{4}_{} 2k_{{\rm F}}^{}}{\alpha^{4}_{} (\hbar \alpha)^2_{}} \quad\quad\quad\quad\quad\;\;\; {\rm in\; region\; VII} \nonumber
\end{eqnarray}
with $C\simeq g_{1}^{2}/(2\pi )^{3}_{}\rho_{0}^{2}\hbar^{2}_{}v$ and $C_{{\rm G}}^{}\simeq C g_{2}^{2}/g_{1}^{2}$, up to numerical prefactors of order one.

\subsection{In-plane vs flexural phonons}

Our analysis allows us to compare the contribution to the resistivity of  in-plane and flexural modes.
In order to quantify the importance of these two contributions at a given temperature, it has to be pointed out that the Bloch-Gr\"uneisen temperatures for in-plane and flexural phonons can be significantly different, in particular in the low-tension regime where out-of-plane modes have a soft quadratic dispersion. For graphene one finds $T_{{\rm BG}}^{(l)}\simeq 50\, \tilde{n}^{1/2}_{}\, {\rm K}$ while (in the absence of tension) $T_{{\rm BG}}^{(h)}\simeq 0.4\, \tilde{n}\, {\rm K}$, with $\tilde{n}=n/10^{12}\, {\rm cm}^{-2}_{}$ the rescaled electron density. 
In the absence of tension at $T=T_{{\rm BG}}^{(l)}$ the estimates above yield a ratio between the scattering rate due to flexural and in-plane phonons of $2.5\,  \tilde{n}_{}^{-1/2}$ and for higher temperatures the flexural modes will be even more dominant.

In practice, at the present electron concentrations for suspended graphene samples, {\em in the absence of tension the flexural-phonon contribution to the resistivity should dominate over the in-plane one at any temperature}, showing a crossover between a $T^{5/2}_{}$ to a $T^{7/2}_{}$ dependence around $T_{{\rm GD}}^{(h)}$, and between $T^{7/2}_{}$ and $T^{2}_{}$ around $T_{{\rm BG}}^{(h)}$. 
The main reason why the effect of flexural modes does not appear in experiments is due to the suppression of the flexural phonon contribution to the resistivity by the sample-specific tension. 
Actually, due to the negative thermal expansion coefficient of graphene, \cite{Lau} tension is itself a temperature dependent quantity.
This suppression leaves the in-plane contribution as the dominant scattering mechanism, yielding a linear-$T$ dependence compatible with experiments. 

When tension is present, in the experimentally relevant regimes $q_*^{}\ll 2k_{{\rm F}}^{}\ll q_T^{(h)}$ and $2k_{{\rm F}}^{}\ll q_*^{}\ll q_T^{(h)}$ the contribution to the resistivity from flexural phonons is approximated by
\begin{equation}\label{RhoFlex}
\Delta\rho_{{\rm flex}}^{}\simeq \frac{h}{e^{2}_{}}\, 10^{-9}_{}\, \frac{\tilde{g}_{1}^{2}}{\tilde{n}+\tilde{\gamma}}\, \tilde{T}^{2}_{}
\end{equation}
with the rescaled tension $\tilde{\gamma}=\gamma /2\cdot 10^{-2}_{}\, {\rm Kg\, s^{-2}_{}}$. The value $\tilde{\gamma }=1$ corresponds to the condition $q_{*}^{}=2k_{{\rm F}}^{}$ at the density $\tilde{n}=1$.
Comparing Eq.\ (\ref{RhoIn}) and (\ref{RhoFlex}) we thus get
\begin{equation}\label{RhoRatio}
\frac{\Delta\rho_{{\rm flex}}^{}}{\Delta\rho_{{\rm in}}^{}}\simeq \frac{1}{20}\cdot\frac{\tilde{T}}{\tilde{n}+\tilde{\gamma}}\; .
\end{equation}
We can then estimate the minimal tension needed to suppress the flexural contribution by imposing the ratio in Eq.\ (\ref{RhoRatio}) to be smaller than unity at room temperature. Considering that in typical current suspended samples $\tilde{n}<1$, this results in a tension $\tilde{\gamma}\simeq 15$, corresponding to a strain of about $10^{-3}_{}$.
These estimates are reliable as long as we do not enter the regime of very high tension $2k_{{\rm F}}^{}\ll q_T^{(h)}\ll q_*^{}$, i.e. for $\tilde{\gamma}< 3\tilde{T}$, which is usually easily fulfilled for $T>T_{{\rm BG}}^{(l)}$.

%that the flexural resistivity becomes smaller than the in-plane one at $T=T_{{\rm BG}}^{({\rm in})}$. From our estimates above, in the regime $2k_{{\rm F}}^{}\ll q_*^{}\ll q_T^{(h)}$, this requires at least $\gamma \gtrsim 3\cdot 10^{-3}_{}\tilde{n}^{1/2}_{}\, {\rm Kg/s^{2}_{}}$, corresponding to a strain of about $10^{-5}_{}\, \tilde{n}^{1/2}_{}$. 
Thus rather weak tension is sufficient to suppress the flexural contribution in favour of the in-plane one. This is true for temperatures up to $\tilde{T}_{c}^{}\simeq 20(\tilde{n}+\tilde{\gamma})$ where the crossover between the in-plane dominated to the flexural-phonon dominated resistivity takes place. 
%In contrast, for $\tilde{T}_{}^{}>\tilde{T}_{c}^{}$ at high enough temperatures, or in samples with reduced tension, one may enter the regime $\tilde{T}> 20(\tilde{n}+\tilde{\gamma})$. 
The observation of the crossover between these two regimes would provide information about the otherwise unknown value of tension in the sample.
This prediction could be tested, for example, in graphene samples mounted on break junctions where tension can be controllably tuned or in flakes clamped on a single side with an STM tip as a drain contact.

\begin{figure}[t]
	\centering
		\includegraphics[width=1.0\columnwidth]{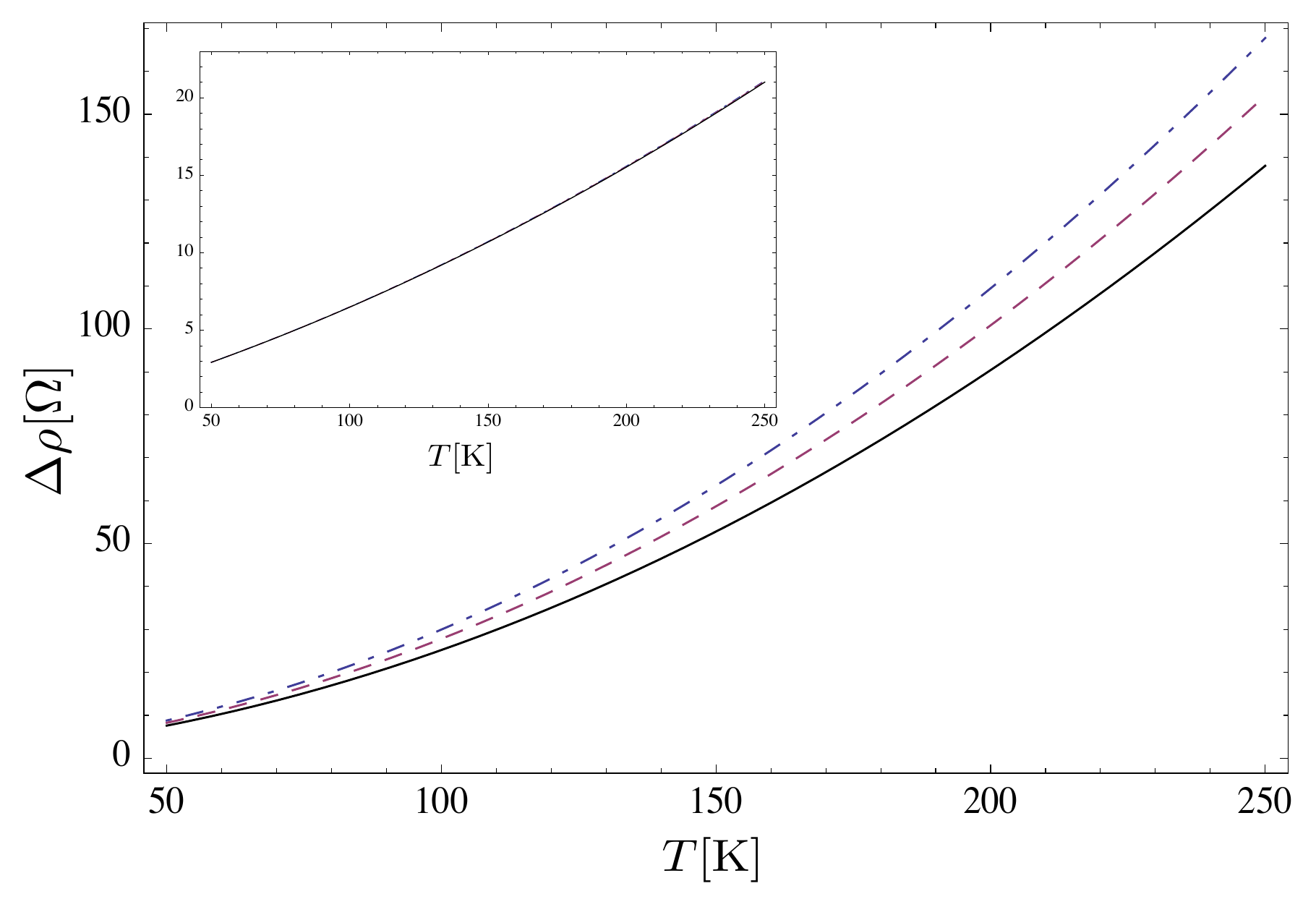}
			\caption{The combined contributions to the resistivity due to in-plane and flexural-phonons $\Delta\rho$ as a function of the temperature $T$ for three different electron densities $\tilde{n}=0.05,\, 0.15,\, 0.3$ (dashed-dotted, dashed and continuous line, respectively). Here we assume a tension $\tilde{\gamma }=1$ and a deformation potential coupling $\tilde{g}_{1}^{}=10$. Inset: same plot as in the main figure, but for stronger tension $\tilde{\gamma }=20$. Notice the almost perfect linear-$T$ scaling, independent of density.
			\label{TempDep}}
\end{figure}

In Fig.\ \ref{TempDep} we plot the temperature-dependent component  of the resistivity $\Delta\rho =\Delta\rho_{{\rm in}}^{}+\Delta\rho_{{\rm flex}}^{}$ according to Eqs.\ (\ref{RhoIn}) and (\ref{RhoFlex}). 
The combined contributions of in-plane and flexural phonons are shown for different densities in the experimentally relevant range. Even in the presence of tension of the order of $\tilde{\gamma }=1$, the  contribution from flexural-phonons can still significantly affect the value of the resistivity. This results in a slight deviation from the purely linear-$T$ dependence due to in-plane modes. In this case, a residual density dependence is observable, stemming from the regime $q_*^{}\ll 2k_{{\rm F}}^{}\ll q_T^{(h)}$, in qualitative agreement with experiments on suspended graphene. \cite{Bolotin}
In contrast, the inset in Fig.\ \ref{TempDep} shows the density-independent $T$-linear resistivity at larger tension ($\tilde{\gamma }=20$).
 
A quantitative understanding of the density dependence observed in experiments would require the knowledge of the sample-specific tension, as well as the inclusion of temperature dependent screening of charged impurities, \cite{Hwang09} of Altshuler-Aronov corrections, \cite{Cheianov} and possibly of further non-intrinsic electron-phonon coupling mechanisms (e.g. capacitive coupling to a back gate as well as buckling). These issues are beyond the scope of the present manuscript.

\section{Conclusions}
\label{sec:Conclusions}
In summary, the interplay between the electronic and phononic degrees of freedom in suspended graphene membranes offers a rich scenario which can be addressed in current transport measurements.

Here we analysed the contribution to the resistivity of suspended graphene due to electron-phonon scattering. We discussed the competition between acoustic in-plane and flexural distortions in various temperature regimes. 
We focused on the intrinsic electron-phonon coupling in graphene due to the interaction of electrons with elastic deformations, taking into account both the (screened) deformation potential and the fictitious (or synthetic) gauge field coupling. Further non-universal coupling mechanisms exist, e.g. via the capacitive interaction of the membrane with a back gate or the breaking of the reflection symmetry (e.g. due to buckling). While we do not discuss these in the present work, they yield a sample-specific linear coupling for flexural phonons which, in presence of tension, would result in a $T$-dependent contribution to the resistivity analogous to in-plane modes.

We find that, for the electron densities achievable in suspended graphene, flexural phonons should dominate over in-plane ones {\em at any temperature in the absence of tension}. In particular, the component of the resistivity due to flexural phonons should show a $T^{5/2}$ dependence for $T\ll T_{{\rm GD}}^{(h)}$ associated with the fictitious gauge field coupling. For $T_{{\rm GD}}^{(h)}\ll T\ll T_{{\rm BG}}^{(h)}$ the $T^{7/2}_{}$ dependence stems from the dominant coupling via the screened deformation potential while for $T\gg T_{{\rm BG}}^{(h)}$ screening is irrelevant and the resistivity scales as $T^2$. 
A sample-specific tension induced by the contacts yields a stiffening of the flexural dispersion, corresponding to a suppressed phonon density of states, while the reflection symmetry with respect to the plane protects their weak quadratic coupling. As a result, tension suppresses the flexural-phonons contribution to the resistivity. We point out that, due to the negative thermal expansion coefficient of graphene, tension in real suspended samples is itself temperature-dependent.

We conclude that it is due to the non-universal tension-induced suppression of the contribution due to flexural-modes that experiments seem to show a resistivity dominated by in-plane phonons alone. 
The latter yield a linear temperature dependent resistivity for $T\gg T_{{\rm BG}}^{(l)}$, in qualitative agreement with experiments. This contribution is however independent of electron density even in the presence of electronic screening and cannot account for the observed density dependence in suspended samples.
A density dependence is induced as long as the flexural phonons contribute significantly to the resistivity which is the case for sufficiently weak tension. 
%We believe that flexural phonons play a role in the understanding of transport measurements in suspended graphene and are possibly involved in the density dependence of the resistivity when the contact-induced tension is not too large.
It would be interesting to probe these issues in samples with a controllable degree of tension, like suspended graphene in break-junctions or in suspended flakes clamped on a single side with an STM-tip as a contact.

\begin{acknowledgments}
Useful discussions with Eduardo Castro and Maresa Rieder are gratefully acknowledged. We acknowledge financial support through SFB 658, SPP 1459 of the Deutsche Forschungsgemeinschaft as well as DIP. One of us (FvO) thanks the Aspen Center for Physics for hospitality during the final stages of this work. 
\end{acknowledgments}


\begin{thebibliography}{99}
\bibitem{Parameters} Typical parameters for graphene are $\kappa =1\,\mathrm{eV}$,  $\mu \simeq 4\lambda =9\,\mathrm{eV}{\rm \AA}^{-2}_{}$, $\rho_{0}^{}\simeq 7.6\cdot 10^{-7}_{}\, {\rm kg}\cdot {\rm m}^{-2}_{}$.
\end{thebibliography}
\end{document}